\documentstyle[prd,aps,epsfig]{revtex}       
\tightenlines
\begin{document}


%
%

\title{Anomalous Quartic Gauge Boson Couplings at Hadron Colliders}

\author{O.\ J.\ P.\ \'Eboli$^1$\footnote{Address after October 2000: Instituto
de F\'{\i}sica da USP, C.P. 66.318, S\~ao Paulo, SP 05389-970, Brazil.}, M.\
C.\ Gonzalez-Garc\'{\i}a$^2$, S.\ M.\ Lietti$^1$, and S.\ F.\ Novaes$^1$ \\}

\address{$^1$Instituto de F\'{\i}sica Te\'orica,
     Universidade  Estadual Paulista, \\
     Rua Pamplona 145, 01405--900, S\~ao Paulo, Brazil \\
     $^2$Instituto de F\'{\i}sica Corpuscular IFIC CSIC, Universidad
     de Valencia,\\
     Edificio Institutos de Paterna, Apartado 2085, 46071 Valencia}

\maketitle


\begin{abstract}

  We analyze the potential of the Fermilab Tevatron and CERN Large Hadron
  Collider (LHC) to study anomalous quartic vector--boson interactions $\gamma
  \gamma Z Z$ and $\gamma \gamma W^+ W^-$. Working in the framework of
  $SU(2)_L \otimes U(1)_Y$ chiral Lagrangians, we study the production of
  photons pairs accompanied by $\ell^+ \ell^-$, $\ell^\pm \nu$, and jet pairs
  to impose bounds on these new couplings, taking into account the unitarity
  constraints.  We compare our findings with the indirect limits coming from
  precision electroweak measurements as well as with presently available
  direct searches at LEPII. We show that the Tevatron Run II can provide
  limits on these quartic limits which are of the same order of magnitude as
  the existing bounds from LEPII searches. LHC will be able to tighten
  considerably the direct constraints on these possible new interactions,
  leading to more stringent limits than the presently available indirect ones.

\end{abstract}

\vskip 2pc


\section{Introduction and Formalism}

Within the framework of the Standard Model (SM), the structure of the
trilinear and quartic vector boson couplings is completely determined by the
$SU(2)_L \times U(1)_Y$ gauge symmetry. The study of these interactions can
either lead to an additional confirmation of the model or give some hint on
the existence of new phenomena at a higher scale \cite{anomalous}. Presently,
the triple gauge--boson couplings are being probed at the Tevatron \cite{teva}
and LEP \cite{lep} through the production of vector boson pairs, however, we
have only started to study directly the quartic gauge--boson couplings
\cite{exp:LEP,lep:ichep}.

It is important to independently measure the trilinear and quartic gauge boson
couplings because there are extensions of the SM \cite{hil:bes} that leave the
trilinear couplings unchanged but do modify the quartic vertices. A simple way
to generate, at tree level, new quartic gauge boson interactions is, for
instance, by the exchange of a heavy boson between vector boson pairs.

The phenomenological studies of the anomalous vertices $\gamma\gamma W^+W^-$
and $\gamma\gamma Z Z$ have already been carried out for $\gamma\gamma$
\cite{bel:bou,ggnos}, $e\gamma$ \cite{our:vvv}, and $e^+ e^-$ \cite{stir} 
colliders.  Some preliminary estimates of the potential of the Tevatron
collider have been also presented in Ref.\ \cite{stir2} where only the effect
on the total cross section for ``neutral'' final states $\gamma W^+ W^-$ and
$\gamma\gamma Z$ were considered while the most promising charged final state
$\gamma \gamma W^\pm$ was not included. In this paper we analyze the potential
of hadron colliders to unravel deviations on the quartic vector boson
couplings by examining the most relevant processes which are the production of
two photons accompanied by a lepton pair, where the fermions are produced by
the decay of either a $W^\pm$ or a $Z^0$ in the anomalous contribution, {\it
i.e.}
\begin{eqnarray} 
p + p~(\bar{p})  &\to& \gamma + \gamma + (W^* \to) \; \ell + \nu  \; ,
\label{ln} \\ 
p + p~(\bar{p}) &\to& \gamma + \gamma + (Z^* \to) \; \ell + \ell   \; ,
\label{ll}
\end{eqnarray} 
as well as the production of photon pairs accompanied by jets
\begin{equation}
p + p \to \gamma + \gamma + j + j 
\label{jj}
\end{equation}
for the LHC.

We carry out a detailed analysis of these reactions taking into account the
full SM background leading to the same final state. We introduce realistic
cuts in order to reduce this background and we include the effect of detector
efficiencies in the evaluation of the attainable limits. We further consider
the energy dependence (form factor) of the anomalous couplings in order to
comply with the unitarity bounds.  Our results show that although the analysis
of Tevatron Run I data can only provide limits on these quartic couplings
which are worse than the existing bounds from LEPII searches, the Tevatron Run
II could yield bounds of the same order of magnitude as the present LEPII
limits. Moreover, the LHC will be able to tighten considerably the direct
constraints on these possible new interactions, giving rise to limits more
stringent than the presently available indirect bounds.

In order to perform a model independent analysis, we use a chiral lagrangian
to parametrize the anomalous $\gamma\gamma W^+W^-$ and $\gamma\gamma Z Z$
interactions \cite{bl}. Assuming that there is no Higgs boson in the low
energy spectrum we employ a nonlinear representation of the spontaneously
broken $SU(2)_L \otimes U(1)_Y$ gauge symmetry. To construct such lagrangian,
it is useful to define the matrix--valued scalar field $\xi(x) = \exp(2iX_a
\varphi^a(x)/v)$, where $X_a$ are the broken generators and $\varphi^a$ are
the Nambu--Goldstone bosons of the global symmetry-breaking pattern $SU(2)_L
\otimes U(1)_Y \to U(1)_{em}$. We denote the unbroken generator by $Q$ and our
conventions are such that $\hbox{Tr} ( X_a X_b) = \frac{1}{2}\delta_{ab}$ and
$\hbox{Tr}(X_a Q) = 0$.

The action of a transformation $G$ of the gauge group $SU(2)_L \otimes U(1)_Y$
on $\xi$ takes the form
\begin{equation}
\xi \to \xi^\prime \;\;\;\;\; \hbox{where} \;\;\;\;\; G \xi = \xi^\prime 
H^\dagger \; .
\end{equation}
$H = \exp(i Q u)$ is defined requiring that $\xi^\prime$ contains only the
broken generators. In order to write the effective lagrangian for the gauge
bosons, it is convenient to introduce the auxiliary quantity
\begin{equation}
{\cal D}_\mu (\xi) \equiv \xi^\dagger \partial_\mu \xi
- i \xi^\dagger (g W^a_\mu T_a + g^\prime B_\mu Y) \xi \; ,
\end{equation}
where $T_a$ and $Y$ are the generators of $SU(2)_L$ and $U(1)_Y$ respectively.

Now we can easily construct fields which have a simple transformation law
under $SU(2)_L \otimes U(1)_Y$:
\begin{eqnarray}
 e {\cal A}_\mu &&\equiv \hbox{Tr} [ Q {\cal D}_\mu(\xi) ]
\;\;\;\;\;\;\;\;\;\;\;\;\;\;\;\;\;\;
 e {\cal A}_\mu \to e {\cal A}_\mu + \partial_\mu u \; ,
\\
\sqrt{g^2+ g^{\prime 2}} {\cal Z}_\mu &&\equiv \hbox{Tr} 
[ X_3 {\cal D}_\mu(\xi) ]
\;\;\;\;\;\;\;\;\;\;\;\;\;\;\;\;\;\;
{\cal Z}_\mu \to {\cal Z}_\mu \; ,
\\
 g {\cal W}^\pm_\mu &&\equiv i \sqrt{2} \hbox{Tr} [ T_\mp {\cal D}_\mu(\xi) ]
\;\;\;\;\;\;\;\;\;\;
{\cal W}^\pm_\mu \to e^{\pm iuQ}~ {\cal W}^\pm_\mu \; ,
\end{eqnarray}
with the standard definition $T_\pm = T_1 \pm i T_2$. Notice that the fields
${\cal A}$, ${\cal Z}$, and ${\cal W}^\pm$ transform only electromagnetically
under $SU(2)_L \otimes U(1)_Y$. Therefore, effective lagrangians must be
invariant exclusively under the unbroken $U(1)_{em}$. Moreover, in the unitary
gauge ($\xi = 1$) we have that ${\cal A} \to A$, ${\cal Z} \to Z$, and ${\cal
W}^\pm \to W^\pm$.

Requiring $C$ and $P$ invariance, the lowest order effective interactions
involving photons is
\begin{eqnarray}
{\cal L}_{\text{eff}} &=& - \frac{\pi \alpha \, \beta_1}{2}\,
F^{\mu\nu}F_{\mu\nu}  {\cal W}^{+ \alpha} {\cal W}^-_\alpha 
- \frac{\pi \alpha \, \beta_2}{4}\,
F^{\mu\nu}F_{\mu\nu}  {\cal Z}^\alpha {\cal Z}_\alpha
\nonumber
\\
&& - \frac{\pi \alpha \, \beta_3}{4}\, 
F^{\mu\alpha}  F_{\mu\beta} ( {\cal W}^{+}_\alpha {\cal W}^{-\beta}
+ {\cal W}^{+}_\beta {\cal W}^{-\alpha} )
- \frac{\pi \alpha \, \beta_4}{4}\, 
F^{\mu\alpha}  F_{\mu\beta} {\cal Z}_\alpha {\cal Z}^{\beta} \; .
\label{l:eff}
\end{eqnarray}
In order to avoid the strong low energy constraints coming from the $\rho$
parameter we impose the custodial $SU(2)$ symmetry which leads to $\beta_1 =
c^2_W \beta_2 = \beta_0$ and $\beta_3 = c^2_W \beta_4 = \beta_c$.  With this
choice ${\cal L}_{eff}$ reduces to the parametrization used in Ref.\
\cite{bel:bou}. In the unitary gauge, (\ref{l:eff}) gives rise to anomalous
$\gamma\gamma ZZ$ and $\gamma\gamma W^+W^-$ vertices which are related by the
custodial symmetry.


\section{Present Constraints: Precision Data, LEPII, and Unitarity Bounds}

The couplings defined in the effective lagrangian (\ref{l:eff}) contribute at
the one--loop level to the $Z$ physics \cite{our:vvv} via oblique corrections
as they modify the $W$, $Z$, and photon two--point functions, and consequently
they can be constrained by precision electroweak data.  We denote the new
contribution to the two--point functions as $\Pi_{VV(0,c)}$ and we take here
the opportunity to update the constraints on $\beta_0$ and $\beta_c$ derived
in Ref.\ \cite{our:vvv}.

It is easy to notice from the structure of the lagrangians that the
contributions to the $W$ and $Z$ self--energies are constant, {\sl i.e.} they
do not depend on the external momentum. Moreover, due to the $SU(2)$ custodial
symmetry they are related by
\begin{equation}
        \Pi_{WW(0,c)} = c^2_w \Pi_{ZZ(0,c)} \; .
\end{equation}
As a consequence the couplings $\beta_0$ and $\beta_c$ do not contribute to
 $T=\Delta \rho$ \cite{stu}. Equivalently their contribution to
 $\sin{\overline{\theta_W}}$ vanishes.  Moreover, the unbroken $U(1)_{em}$
 symmetry constrains the photon self-energy contribution to be of the form
\begin{equation}
        \Pi_{\gamma\gamma(0,c)}(q^2) = q^2 \Pi^\prime_{\gamma\gamma(0,c)} \;,
\end{equation}
where for the anomalous interactions (\ref{l:eff}) $\Pi^\prime_{\gamma
\gamma(0,c)}$ is a constant. This also implies that these anomalous
interactions do not modify the running of the electromagnetic
coupling. However, both interactions give rise to corrections to $\Delta r$
or, equivalently, to the $S$ and $U$ parameters \cite{stu}.

Following the standard procedure, we evaluated the vector boson two--point
functions using dimensional regularization and subsequently kept only the
leading non--analytic contributions from the loop diagrams to constrain the
new interactions --- that is, we maintained only the logarithmic terms,
dropping all others. The contributions that are relevant for our analysis are
easily obtained by the substitution
\[
\frac{2}{4-d} \rightarrow {\rm{log}}\;\frac{\Lambda^2}{\mu^2}\; ,
\]
where $\Lambda$ is the energy scale which characterizes the appearance
of new physics, and $\mu$ is the scale in the process, which we take
to be $M_W$. After this procedure we obtain that 
\begin{eqnarray}
\alpha S = &&-4 s_W^2 c_W^2 \Pi_{\gamma \gamma}'
\nonumber
\\
 = &&-4 s_W^2 c_W^2 \left \{ \frac{\alpha \, \beta_0  M_W^2}{4 \pi}
\left [ -\left(1 + \frac{1}{2 c_W^4}\right)  +\frac{3}{2} 
\ln\left(\frac{\Lambda^2}{M_W^2}\right)
       +\frac{3 }{4 c_W^4}
         \ln\left(\frac{\Lambda^2 c_W^2}{M_W^2}\right) \right ] \right.
\nonumber \\
&&+ \left. \frac{\alpha \, \beta_c  M_W^2}{64 \pi}
\left [ -\left(1 + \frac{1}{2 c_W^4}\right) +6 
\ln\left(\frac{\Lambda^2}{M_W^2}\right) +\frac{3}{c_W^4}
         \ln\left(\frac{\Lambda^2 c_W^2}{M_W^2}\right) \right ]. \right \}
\; ,
\\
\alpha U = && \frac{s_w^2}{c_W^2} S \; .
\end{eqnarray}
The allowed ranges of $S$ and $U$ depend on the SM parameters.  As an
illustration of the size of the bounds, we take that for the Higgs boson mass
of $M_H=300$ GeV, the 95\%CL limits on $S$ and $U$ are $0.34 \leq S \leq 0.02$
and $-0.13 \leq U \leq 0.37$ \cite{pdg}. These bounds can then be translated
into the 95\% CL limits on $\beta_0$ and $\beta_c$ presented in Table
\ref{LFOP}.

The LEP collaborations have directly probed anomalous quartic couplings
involving photons.  L3 and OPAL have searched for their effects in the
reactions $e^+ e^- \to W^+ W^- \gamma$, $Z \gamma \gamma$, and $\nu \bar{\nu}
\gamma \gamma$, while the ALEPH collaboration has reported results only on the
last reaction \cite{exp:LEP,lep:ichep}.  The combined results for all these
searches lead to the following 95\% CL direct limits on the quartic vertices
\cite{lep:ichep}
\begin{eqnarray}
-4.9\times 10^{-3} \hbox{ GeV}^{-2} ~<~ &\beta_0& ~<~ 5.6\times 10^{-3} 
\hbox{ GeV}^{-2}\; ,
\\
-5.4\times 10^{-3} \hbox{ GeV}^{-2} ~<~ &\beta_c&
~<~ 9.8\times 10^{-3} \hbox{ GeV}^{-2}\; .
\end{eqnarray}

Another way to constrain the couplings in (\ref{l:eff}) is to notice that this
effective lagrangian leads to tree-level unitarity violation in $2\to2$
processes at high energies. In order to extract the unitarity bounds on the
anomalous interactions we evaluated the partial wave helicity amplitudes
($\tilde{a}^j_{\nu\mu}$) for the inelastic scattering $\gamma(\lambda_1)
\gamma(\lambda_2) \to V(\lambda_3) V(\lambda_4)$, with $V=Z$ and $W^\pm$; see
Table \ref{a0numu}. Unitarity requires that \cite{bz}
\begin{equation}
\beta_V  \sum_\nu | \tilde{a}^j_{\nu \mu}|^2
\le \frac{1}{4} \; ,
\label{dieter}
\end{equation}
where $\beta_V$ is the velocity of the final state boson in the
center--of--mass frame.  For the anomalous interactions (\ref{l:eff}), the
most restrictive bounds come from the $J=0$ partial wave, which read
\begin{eqnarray}
&&\left(\frac{\alpha \beta s}{16} \right)^2 \left(1-\frac{4
M_W^2}{s}\right)^{1/2}\left(3 - \frac{s}{M_W^2} + \frac{s^2}{4
M_W^4}\right) \le  N \hbox{   for   } V= W \; ,
\\
&&\left(\frac{\alpha \beta s}{16 c_W^2} \right)^2 \left(1-\frac{4
M_Z^2}{s}\right)^{1/2}\left(3 - \frac{s}{M_Z^2} + \frac{s^2}{4
M_Z^4}\right) \le  N \hbox{   for   } V=Z\; ,
\end{eqnarray}
where $\beta = \beta_0$ or $\beta_c$ and $N = 1/4 \; (4)$ for $\beta_0$
($\beta_c$). For instance, unitarity is violated for $\gamma\gamma$ invariant
masses above 240 GeV for $\beta_0= 5.6\times 10^{-3}$ GeV$^{-2}$ (one of the
present LEP bounds).

These unitarity constraints are of relevance when extracting the bounds on the
anomalous couplings at hadron colliders since it is possible to obtain large
parton--parton center--of--mass energies, and consequently have a large
unitarity violation.  The standard procedure to avoid this unphysical behavior
of the subprocess cross section and to obtain meaningful limits is to multiply
the anomalous couplings by a form factor
\begin{equation}
\beta_{0,c} \longrightarrow \left(1 +
\frac{M_{\gamma\gamma}^2}{\Lambda^2}\right)^{-n} \times \beta_{0,c}
\; ,
\label{ff}
\end{equation}
where $M_{\gamma\gamma}$ is the invariant mass of the photon pair.  Of course
using this procedure the limits become dependent on the exponent $n$ and the
scale $\Lambda$ which is not longer factorizable.  In our calculations, we
conservatively choose $n=5$ and $\Lambda$ = 0.5 TeV for the Tevatron and
$\Lambda$ = 0.5 (2.5) TeV for the LHC. In the case of $e^+e^-$ colliders the
center--of--mass energy is fixed and the introduction of the form factor
(\ref{ff}) is basically equivalent to a rescaling of the anomalous couplings
$\beta_{0,c}$, therefore we should perform this rescaling when comparing 
results obtained at hadron and $e^+e^-$ colliders. For example, for our 
choice of $n$ and $\Lambda$ the LEP limits should be weakened by a factor 
$\simeq 1.6$.

\begin{figure}
\protect
\centerline{\mbox{\epsfig{file=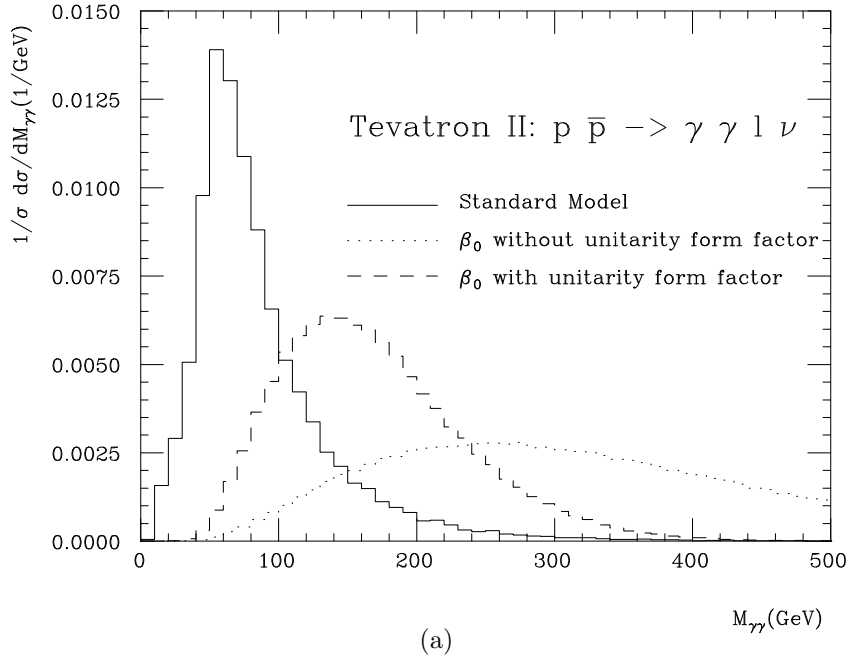,angle=90,width=0.75\textwidth}}}
\vskip -1cm 
\centerline{\mbox{(a)}}
\centerline{\mbox{\epsfig{file=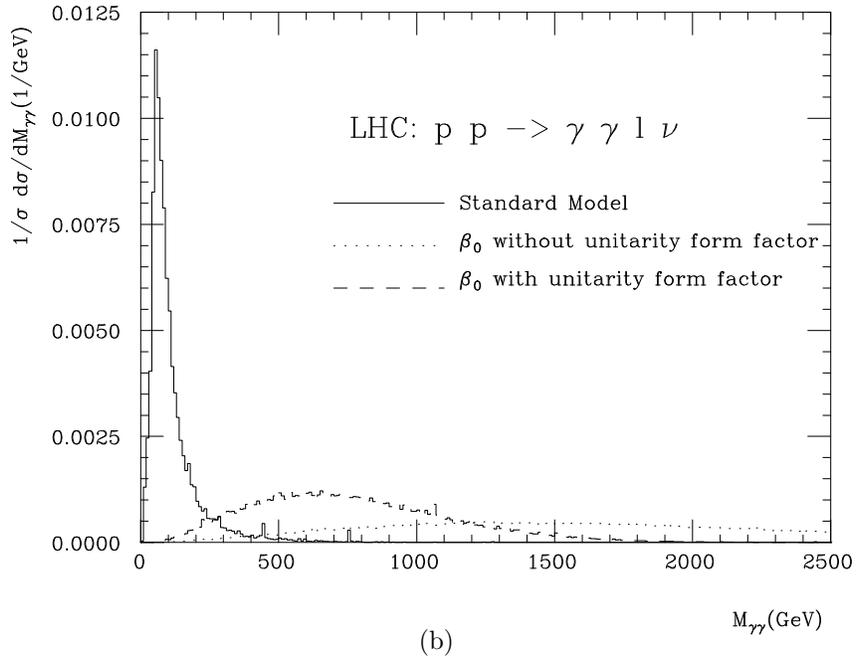,angle=90,width=0.75\textwidth}}}
\vskip -1cm 
\centerline{\mbox{(b)}}
\vskip 1cm 
\caption{Normalized invariant mass distribution of the $\gamma\gamma$
pair for the reaction $p + p~(\bar{p}) \to \gamma + \gamma + (W^* \to) \;
\ell + \nu$ at Tevatron Run II (a) and LHC (b). The solid histogram
represents the SM contribution while dashed (dotted) histograms is the
anomalous $\beta_0$ contribution with (without) unitarity form factor.  We
chose $n=5$ and $\Lambda = 0.5$ (2.5) TeV for the Tevatron (LHC).}
\label{m:gg}
\end{figure}

The dynamical effect of the above form factor can be seen in Figure \ref{m:gg}
where we present the normalized invariant mass distribution of the
$\gamma\gamma$ pair for the process (\ref{ln}) at the Tevatron Run II and LHC,
assuming that only $\beta_0$ contributes.  As expected, the form factor
reduces the number of photon pairs with high invariant mass.  Similar behavior
is obtained for reaction (\ref{ll}) and for the anomalous $\beta_c$
contribution.

\section{Signals at Hadron Colliders}

In this work we studied the reactions (\ref{ln}) and (\ref{ll}) for the
Tevatron and LHC, that is, the associated production of a photon pair and a
$W^*$ or $Z^*$ which decay leptonically, as well as the process (\ref{jj})
only for the LHC since the Tevatron center--of--mass energy is too low for
this process to be of any significance. Process (\ref{ln}) can be used to
study the $\gamma\gamma W^+ W^-$ vertex while process (\ref{ll}) probes the
$\gamma\gamma ZZ$ interaction and the reaction (\ref{jj}) receives
contributions from $\gamma\gamma W^+ W^-$ and $\gamma\gamma ZZ$.  We evaluated
numerically the helicity amplitudes of all the SM subprocesses leading to the
$\gamma\gamma\ell^\pm\nu$, $\gamma\gamma\ell^+\ell^-$, and $\gamma\gamma jj$ 
final states where $j$ can be either a gluon, a quark or an antiquark. 
The SM amplitudes were generated using Madgraph
\cite{mad} in the framework of Helas \cite{helas} routines.  The anomalous
interactions arising from the Lagrangian (\ref{l:eff}) were implemented as
subroutines and were included accordingly. We consistently took into account
the effect of all interferences between the anomalous and the SM amplitudes,
and did not use the narrow--width approximation for the vector boson
propagators.

In the case of the Tevatron collider, we considered the parameters of the Run
I, {\it i.e.}  $\sqrt{s} = 1.8$ TeV and an integrated luminosity of 100
pb$^{-1}$. We also investigated the reach of the Tevatron Run II assuming
$\sqrt{s} = 2$ TeV and an integrated luminosity of $2 \times 10^3$
pb$^{-1}$. For the LHC, we took a center--of--mass energy of 14 TeV
and a luminosity of 10$^5$ pb$^{-1}$. In our calculations we used the MRS (G)
\cite{mrs} set of proton structure functions with the factorization scale 
$Q^2 = \hat{s}$.

We started our analysis of the processes (\ref{ln}) and (\ref{ll}) imposing a
minimal set of cuts to guarantee that the photons and charged leptons are
detected and isolated from each other:
\begin{eqnarray}
p_{T}^{(\ell,\nu)}& \geq & 20~~(25) \text{ GeV for } \ell = e~~ (\mu)
\; , \nonumber \\
E_{T}^{\gamma}& \geq & 20 \text{ GeV }
\; , \nonumber \\
|\eta_{\gamma,e}| & \leq & 2.5
\; , 
\label{cuts} \\
|\eta_{\mu}| & \leq & 1.0
\; ,
\nonumber \\
\Delta R_{ij} & \geq & 0.4 \; ,
 \nonumber
\end{eqnarray}
where $i$ and $j$ stand for the final photons and charged leptons.
For the $\gamma \gamma \ell \nu$ final state, we also imposed a cut 
the transverse mass of the $\ell \nu$ pair
($M^{\ell \nu}_T$) 
\begin{equation}
65 \text{ GeV } \leq  M^{\ell \nu}_T  \leq 100 \text{ GeV }  \; . 
\label{cuts1}
\end{equation}
In the case of $\gamma \gamma \ell^+ \ell^-$ production, we required the tag
of a $Z$ decaying leptonically imposing that
\begin{eqnarray}
75 \text{ GeV } \leq  M^{\ell \ell} \leq 105 \text{ GeV}   \; ,  
\label{cuts2}
\end{eqnarray}
where $M^{\ell \ell}$ is the invariant mass of the lepton pair.  In our
calculations, we have also taken into account the detection efficiency of the
final state particles.  We assumed a 85\% detection efficiency of isolated
photons, electrons, and muons.  Therefore, the efficiency for reconstructing
the final state $\gamma \gamma \ell \nu$ is 61\% while the efficiency for
$\gamma \gamma \ell^+ \ell^- $ is $52$\%.

Considering the cuts (\ref{cuts}), (\ref{cuts1}), and (\ref{cuts2}), and the
detection efficiencies discussed above, the SM prediction for the cross
sections and expected number of events of the processes (\ref{ln}) and
(\ref{ll}) are presented in Table \ref{SMEv}. As we can see, the above basic
cuts are enough to eliminate the SM background at the Tevatron Run I, however,
further cuts are needed to control the background at the Tevatron Run II 
and LHC.

\begin{figure}
\protect
\centerline{\mbox{\epsfig{file=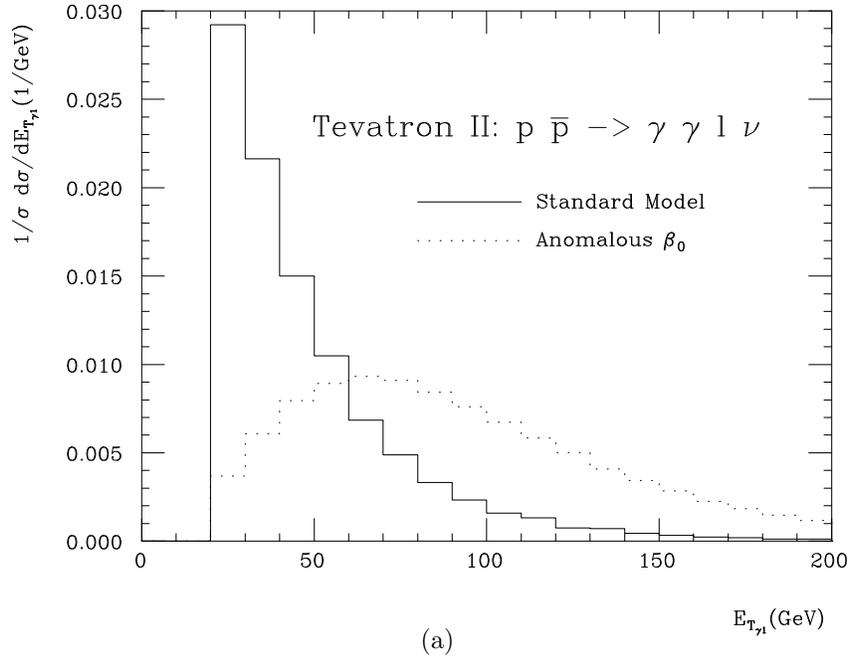,angle=90,width=0.75\textwidth}}}
\vskip -1cm 
\centerline{\mbox{(a)}}
\centerline{\mbox{\epsfig{file=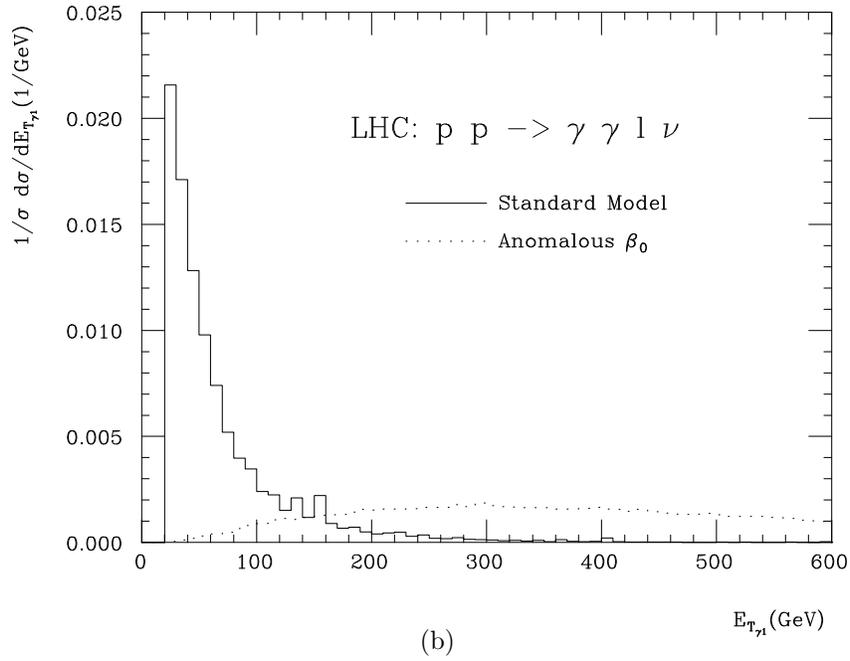,angle=90,width=0.75\textwidth}}}
\vskip -1cm 
\centerline{\mbox{(b)}}
\vskip 1cm
\caption{Normalized transverse energy distribution of the most
energetic photon for the reaction (\protect{\ref{ln}}) at Tevatron Run II (a)
and LHC (b). The solid histogram represents the SM contribution while the
dotted one is the anomalous $\beta_0$ contribution.}
\label{ET}
\end{figure}

In order to reduce the SM background for the Tevatron Run II and LHC, we
analyzed a few kinematical distributions. The most significant difference
between the SM and anomalous predictions appears in the transverse energy of
the photons, which is shown in Figure \ref{ET} for the reaction (\ref{ln}) and
$\beta_0 \ne 0$.  Similar behavior is obtained for the reaction (\ref{ll}) and
for the anomalous $\beta_c$ contribution. Therefore, we tightened the cut on
the transverse energy of the final photons, as suggested by Fig.\ \ref{ET}, to
enhance the significance of the anomalous contribution.
\begin{eqnarray}
E_{T}^{{\gamma_{1(2)}}} & \geq & 75~~(50) \text{ GeV for Tevatron Run II and}
\label{new:cuts} \\
E_{T}^{{\gamma_{1(2)}}} & \geq & 200~~(100) \text{ GeV for LHC}\; .
\nonumber 
\end{eqnarray}
The effect of these cuts can be seen in Table \ref{SMEv} where we display the
new cross sections and expected number of events in parenthesis. As we can
see, no SM event is expected at the Tevatron Run II after this new cut, while
very few events survive at the LHC.

We parametrized the cross sections for processes (\ref{ln}) and (\ref{ll})
after cuts (\ref{cuts})--(\ref{new:cuts}) as
\begin{equation}
\sigma \equiv \sigma_{\text{sm}} + \beta~ \sigma_{\text{inter}} + \beta^2~ 
\sigma_{\text{ano}} \; ,  
\label{crosssection}
\end{equation}
where $\sigma_{\text{sm}}$, $\sigma_{\text{inter}}$, and $\sigma_{\text{ano}}$
are, respectively, the SM cross section, interference between the SM
and the anomalous contribution and the pure anomalous cross section. 
$\beta$ stands for $\beta_0$ or $\beta_c$.  The results for 
$\sigma_{\text{sm}}$, $\sigma_{\text{inter}}$, and
$\sigma_{\text{ano}}$ are presented in Table
\ref{tabcrosssection}.

\begin{figure}
\protect
\centerline{\mbox{\epsfig{file=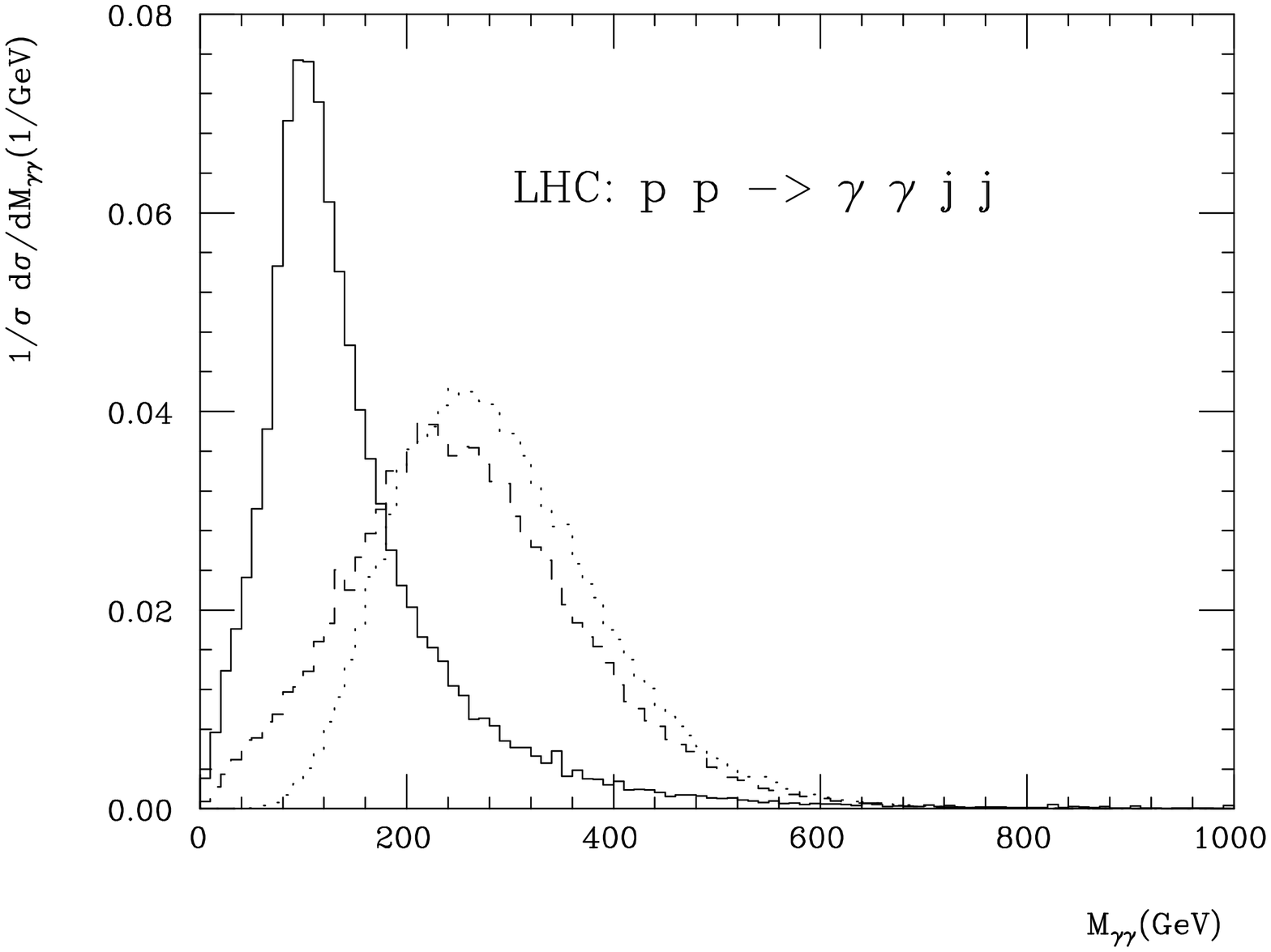,angle=90,width=0.75\textwidth}}}
\vskip -1cm
\centerline{\mbox{(a)}}
\centerline{\mbox{\epsfig{file=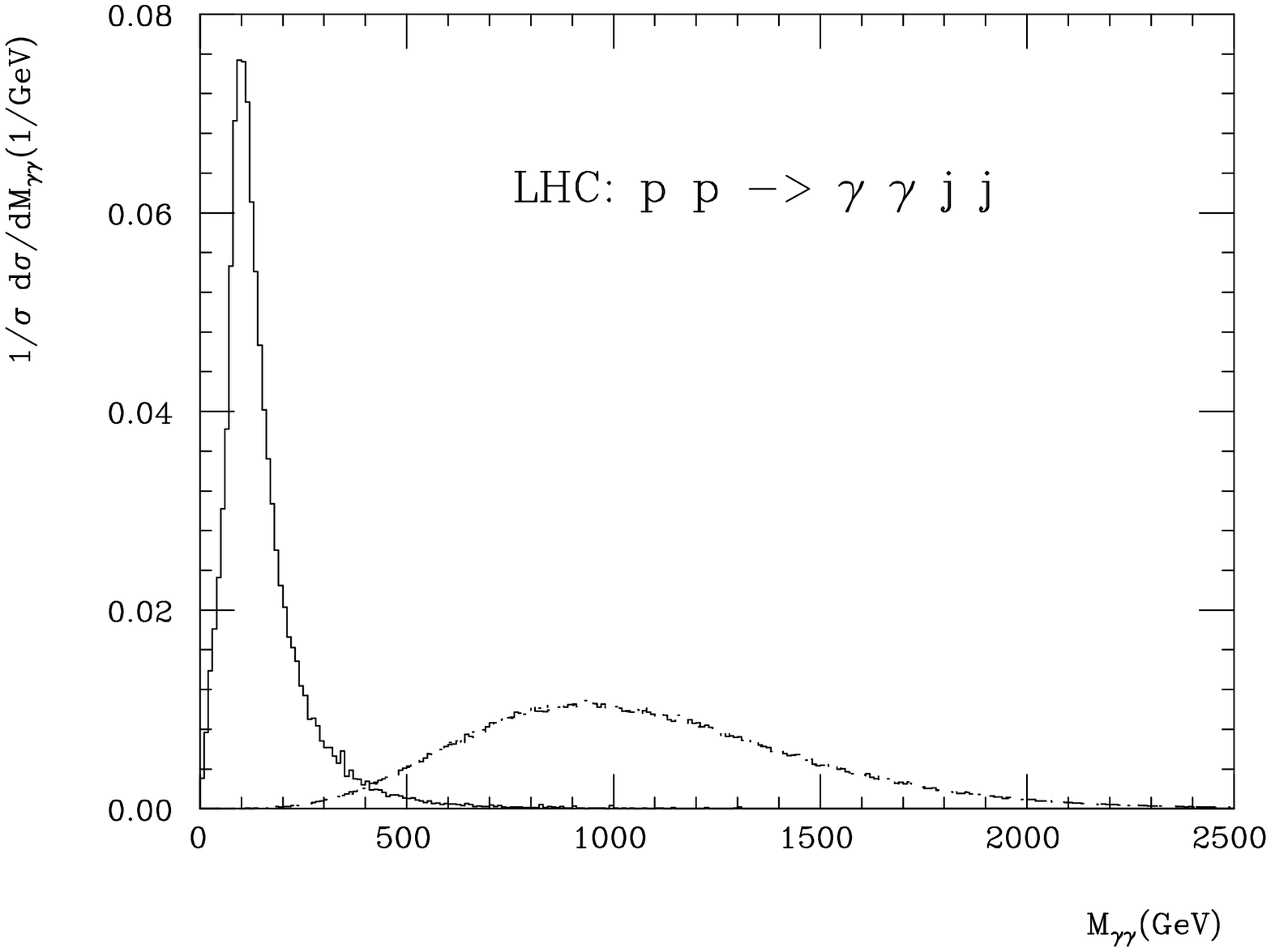,angle=90,width=0.75\textwidth}}}
\vskip -1cm
\centerline{\mbox{(b)}}
\vskip 1cm
\caption{Normalized invariant mass distribution of the $\gamma\gamma$
pair for the reaction $p + p \to \gamma + \gamma + jet + jet $ at
LHC. The solid histogram
represents the SM contribution while dotted (dashed) histograms is the
anomalous $\beta_0$ ($\beta_c$) contribution with unitarity form factor.  We
chose $n=5$ and (a) $\Lambda = 0.5$ TeV and (b) $\Lambda = 2.5$ TeV.}
\label{IM}
\end{figure}

Process (\ref{jj}) receives contributions from $W^*$ and $Z^*$ productions
and their subsequent decay into jets, as well as from vector boson fusion (VBF)
\begin{equation}
p + p  \to q + q + (W^* + W^* \mbox{ or } Z^* + Z^*) \to q + q + \gamma + 
\gamma \; .
\label{vbf}
\end{equation}     
The signal for hadronic decays of $W$'s and $Z$'s is immersed in a huge QCD
background. Therefore, we tuned our cuts in order to extract the VBF
production of photon pairs since it presents two very energetic forward jets
that can be used to efficiently tag the events. In our analyses, we required
that the photons satisfy
\begin{eqnarray}
E_{T}^{\gamma_{1(2)}}       & > & 50~ (25)  \; \text{GeV} \; , 
\label{kuts2a}\\
|\eta_{\gamma_{(1,2)}}|      & < & 5.0  \; , \nonumber 
\end{eqnarray}
while the jets should comply with
\begin{eqnarray}
p_{T}^{j_{1(2)}}       & > & 40~ (20)  \; \text{GeV} \; , \nonumber \\
|\eta_{j_{(1,2)}}|      & < & 5.0  \; , \nonumber \\
|\eta_{j_{1}} - \eta_{j_{2}}| & > & 4.4 \; , \nonumber \\
\eta_{j_{1}} . \eta_{j_{2}} & < 0 & \; , 
\label{kuts2b} \\
\text{min}\{\eta_{j_{1}}, \eta_{j_{1}} \}  + 0.7 <
& \eta_{\gamma_{(1,2)}} &
< \text{max}\{ \eta_{j_{1}}, \eta_{j_{1}} \} - 0.7 \; , \nonumber \\
\Delta R_{jj}           & > & 0.7  \; , \nonumber \\
\Delta R_{j\gamma}      & > & 0.7  \; .
\nonumber
\end{eqnarray}
Assuming a 85\% detection efficiency of isolated photons, the efficiency for
reconstructing the final state $jet + jet + \gamma + \gamma$ is 72\%.  Table
\ref{SMEv} also contains the SM cross section for the VBF production of photon
pair taking into account the above cuts. As we can see, the VBF reaction
possesses a much higher statistics than the production of photon pairs
associated to leptons. In order to enhance the VBF signal for the anomalous
couplings we studied a few kinematical distributions and found out that the
most significant difference between the signal and SM background occurs in the
diphoton invariant mass spectrum; see Fig.\ \ref{IM}. Thusly, we imposed the
following additional cuts
\begin{eqnarray}
200~ (400) \text{ GeV } \leq M_{\gamma \gamma} \leq  700~ (2500)
\text{ GeV for } \Lambda = 500~(2500) \text{ GeV.}
\label{new:kuts}
\end{eqnarray}
This cut reduces the SM background cross section by a factor of at least 5;
see Table \ref{SMEv} where we also present the signal cross section after cuts
for $\Lambda=$ 500 and 2500 GeV. The results for $\sigma_{\text{sm}}$ and
$\sigma_{\text{ano}}$ of Eq.\ (\ref{crosssection}) are presented in Table
\ref{tabcrosssection}. Since the interference between the SM and the anomalous
contribution is negligible in this case, we do not present the results for
$\sigma_{\text{inter}}$.

Taking into account the integrated luminosities of the Tevatron and LHC and
the results shown in Table \ref{tabcrosssection}, we evaluated the potential
95\% CL limits on $\beta_0$ and $\beta_c$ in the case that there is no
deviation from the SM predictions; see Table \ref{LFTaLHC}. We also exhibit in
this table our choice for the scale $\Lambda$ appearing in the form
factor. Therefore, at the Tevatron, the most restrictive constraints are
obtained from the reaction (\ref{ln}) for $\beta_0$ and $\beta_c$. Combining
both reactions we are able to impose a 95\% CL limit $|\beta_{0,c}| \lesssim
1.5 \times 10^{-2}$ GeV$^{-2}$ at the Tevatron Run II, which is of the same
order of the direct bounds coming from LEPII. On the other hand, the most
stringent limits at the LHC will come from the photon pair production via VBF,
whose bounds are a factor of 5--10 stronger than the ones coming from the
reactions (\ref{ln}) and (\ref{ll}).  This general statement does not seem to
apply for the limits on $\beta_c$ with $\Lambda=$ 500 GeV, which is more
strongly constrained by the process (\ref{ln}). This is not surprising
because, for the reactions (\ref{ln}) and (\ref{ll}), the set of cuts
(\ref{new:cuts}) leave the $\beta_c$ signal practically unaffected, i.e. this
set of general cuts is particularly optimum for this coupling and
reactions. This is also there reason why the derived limits on $\beta_c$ are
better than the limits for $\beta_0$ only for this case.

\section{Summary and Conclusions}

We are just beginning to test the SM predictions for the quartic vector boson
interactions. Due to the limited available center--of--mass energy, the first
couplings to be studied contain two photons, and just at the LHC and the NLC
we will be able to probe $VVVV$ ($V=W$ or $Z$) vertices \cite{vvvv}.  In this
work we analyzed the production of photon pairs in association with $\ell^\pm
\nu$, $\ell^+\ell^-$, or $jj$ in hadron colliders. These processes violate
unitarity at high energy, therefore, we cutoff the growth of the subprocess
cross section via the introduction of form factors which enforce unitarity and
render the calculation meaningful.

We showed that the study of the processes (\ref{ln}) and (\ref{ll}) at
Tevatron Run I lead to constraints on the quartic anomalous couplings that are
a factor of four weaker than the presently available bounds derived from LEPII
data. On the other hand, the Tevatron Run II has the potential to probe the
quartic anomalous interactions at the same level of LEPII.  An important
improvement on the bounds on the genuine quartic couplings will be obtained at
the LHC collider where, for $\Lambda = 2.5$ TeV, a limit of $|\beta_{0,c}|
\lesssim 10^{-5}$ GeV$^{-2}$ will be reached. Therefore, the direct limits on
the anomalous interaction steaming from LHC will be stronger than the ones
coming from the precise measurements at the $Z$ pole.  It is interesting to
notice that the LHC will lead to limits that are similar to the ones
attainable at an $e^+e^-$ collider operating at $\sqrt{s} = 500 $ GeV with a
luminosity of 300 pb$^{-1}$, which are $|\beta_{0,c}| \lesssim 3 \times
10^{-5}$ GeV$^{-2}$ \cite{stir}.

In conclusion, the LHC will be able to impose quite important limits on
genuine quartic couplings studying the $\gamma\gamma\ell^+\ell^-$,
$\gamma\gamma\ell\nu$, and $\gamma\gamma j j$ productions.


\acknowledgments

This work was supported by Conselho Nacional de Desenvolvimento
Cient\'{\i}fico e Tecnol\'ogico (CNPq), by Funda\c{c}\~ao de Amparo \`a
Pesquisa do Estado de S\~ao Paulo (FAPESP), by Programa de Apoio a N\'ucleos
de Excel\^encia (PRONEX), by the spanish DGICYT under grants PB98-0693 and
PB97-1261, by the Generalitat Valenciana under grant GV99-3-1-01, and by the
TMR network grant ERBFMRXCT960090 of the European Union.


\newpage 

\widetext

\begin{table}
\begin{tabular}{||c||c||c||c||}
$\Lambda$ (TeV) & Parameter & $\beta_0$ (GeV$^{-2}$) & $\beta_c$
(GeV$^{-2}$)\\
\hline
\hline
0.5 & S & 
( $-$0.09 , 1.5 )$\times 10^{-4}$ & ( $-$0.29 , 4.9 )$\times 10^{-4}$ \\
& U & 
( $-$5.4 , 1.9 )$\times 10^{-4}$ & ( $-$18. , 6.2 )$\times 10^{-4}$ \\ 
\hline 
2.5  & S & 
( $-$0.04 , 0.69 )$\times 10^{-4}$ & ( $-$0.15 , 2.5 )$\times 10^{-4}$ \\
& U & 
( $-$2.5 , 0.88 )$\times 10^{-4}$ & ( $-$9.1 , 3.2)$\times 10^{-4}$ 
\end{tabular}
\medskip
\caption{
95\% CL limits on $\beta_o$ and $\beta_c$ steaming from oblique parameters $S$
and $U$.
}
\label{LFOP}
\end{table}


\begin{table}
\begin{tabular}{||c||c||}
($\lambda_1,\lambda_2,\lambda_3,\lambda_4$) & $\tilde{a}^0_{\nu \mu}$ \\
\hline
\hline
($++++$) or ($--++$) & 
$\left(\frac{\alpha s}{16 n_V}\right) \beta$\\
\hline
($++--$) or ($----$) & 
$\left(\frac{\alpha s}{16 n_V}\right) \beta$\\ 
\hline
($++00$) or ($--00$) & 
$\left(1-\frac{s}{2M_V^2}\right) \left(\frac{\alpha s}{16 n_V}\right) \beta$
\end{tabular}
\medskip
\caption{
  $\tilde{a}^0_{\nu\mu}$ for the reactions 
  $\gamma(\lambda_1) \gamma(\lambda_2) \to V(\lambda_3) V(\lambda_4)$, 
  with $V=Z$ and $W^\pm$, where
  $\mu=\lambda_1-\lambda_2$ and $\nu=\lambda_3-\lambda_4$.
  $\beta$ stands for $\beta_0$ or $\beta_c$ and 
  $n_{W^\pm} = 1~(4)$ for $\beta_0~(\beta_c)$ and
  $n_{Z} = c_W^2~(4c_W^2)$ for $\beta_0~(\beta_c)$.  }
\label{a0numu}
\end{table}


\begin{table}
\begin{tabular}{||c||c||c||c||}
Collider & Process & Cross Section (pb) & Number of Events \\ 
\hline
\hline
Tevatron I & $p \bar{p} \to l^\pm \nu_{l^\pm} \gamma \gamma$ & 
1.93 $\times 10^{-4}$ & 1.93$\times 10^{-2}$ \\
           & $p \bar{p} \to l^+ l^- \gamma \gamma$ & 
1.58 $\times 10^{-4}$ & 1.58$\times 10^{-2}$ \\
\hline
Tevatron II & $p \bar{p} \to l^\pm \nu_{l^\pm} \gamma \gamma$ & 
2.13 $\times 10^{-4}$ (7.89 $\times 10^{-6}$) & 
0.43 (1.58  $\times 10^{-2}$) \\
            & $p \bar{p} \to l^+ l^- \gamma \gamma$ & 
1.77 $\times 10^{-4}$ (5.90 $\times 10^{-6}$) & 
0.35 (1.18  $\times 10^{-2}$) \\
\hline
LHC & $p p \to l^\pm \nu_{l^\pm} \gamma \gamma$ & 
1.08 $\times 10^{-3}$ (1.32 $\times 10^{-5}$) & 
108 (1.3) 
\\
    & $p p \to l^+ l^- \gamma \gamma$ & 
6.45 $\times 10^{-4}$ (4.25 $\times 10^{-6}$) & 
65 (0.43) 
\\
    & $pp \to j j \gamma \gamma$ & 3.19$\times 10^{-2}$~ (6.28$\times
10^{-3}$) [1.12$\times 10^{3}$] & 
3190 ~(628) ~[112]

\end{tabular}
\medskip
\caption{ SM cross sections after the cuts.  We applied the cuts
(\protect{\ref{cuts}})--(\protect{\ref{cuts2}}) to the $\ell \nu \gamma\gamma$
and $\ell^+ \ell^- \gamma \gamma$ processes while we used the cuts
(\protect{\ref{cuts}}) and (\protect{\ref{kuts2a}},\protect{\ref{kuts2b}}) 
to the $\gamma \gamma j j $ final state.  We present between
parenthesis the Tevatron II results after we
included the additional cut (\protect{\ref{new:cuts}}) for $\ell \nu
\gamma\gamma$ and $\ell^+ \ell^- \gamma \gamma$ productions. In the case
of $jj\gamma\gamma$ production at LHC, we exhibit between
parenthesis (brackets) the results after cuts 
(\protect\ref{new:kuts}) 
for $\Lambda = 0.5$ (2.5) TeV.}
\label{SMEv}
\end{table}


\begin{table}
\begin{tabular}{||c||c||c||c||c||}
Collider & Process & $\sigma_{\text{sm}}$ (pb) & 
$\sigma_{\text{inter}}$ (pb $\times$ GeV$^{2}$) for $\beta_0$ ($\beta_c$) & 
$\sigma_{\text{ano}}$ (pb $\times$ GeV$^{4}$) for $\beta_0$ ($\beta_c$) \\ 
\hline
\hline 
Tevatron I & $p \bar{p} \to l^\pm \nu_{l^\pm} \gamma \gamma$ & 1.93
$\times 10^{-4}$ & 5.09(2.58)$\times 10^{-3}$ & 15.0(5.50)\\ 
$\Lambda$=0.5 TeV & $p \bar{p} \to l^+ l^- \gamma \gamma$ & 
1.58 $\times 10^{-4}$ & 7.18(1.22)$\times 10^{-3}$ & 3.63(1.37)\\ 
\hline 
Tevatron II & $p \bar{p} \to
l^\pm \nu_{l^\pm} \gamma \gamma$ & 7.89 $\times 10^{-6}$ & 1.20(1.03)$\times
10^{-3}$ & 6.21(2.92)\\ 
$\Lambda$=0.5 TeV & $p \bar{p} \to l^+ l^- \gamma
\gamma$ & 5.90 $\times 10^{-6}$ & 1.38(0.36)$\times 10^{-3}$ & 1.78(0.86)\\
\hline 
LHC & $p p \to l^\pm \nu_{l^\pm} \gamma \gamma$ & 1.32$\times 10^{-5}$
& 3.13(3.97)$\times 10^{-4}$ & 6.79(59.2)\\ 
$\Lambda$=0.5 TeV & $p p \to l^+
l^- \gamma \gamma$ & 4.25$\times 10^{-6}$ & 6.06(0.49)$\times 10^{-4}$ &
4.82(18.5)\\ 
& $p p \to j j \gamma \gamma$ & 6.28$\times 10^{-3}$ & --- &
1.02$\times 10^4$ (7.56$\times 10^2$)\\ 
\hline 
LHC & $p p \to l^\pm
\nu_{l^\pm} \gamma \gamma$ & 1.32$\times 10^{-5}$ & 1.17(22.4)$\times 10^{-3}$
& 5570(2900)\\ 
$\Lambda$=2.5 TeV & $p p \to l^+ l^- \gamma \gamma$ &
4.25$\times 10^{-6}$ & 1.15(1.08)$\times 10^{-2}$ & 3980(1390)\\ 
& $p p \to j j \gamma \gamma$ & 1.12$\times 10^{-3}$ & --- & 
1.07$\times 10^7$ (7.34$\times 10^5$)
\end{tabular}
\medskip
\caption{Results for $\sigma_{\text{sm}}$, $\sigma_{\text{inter}}$ and
$\sigma_{\text{ano}}$; see
Eq.~(\protect{\ref{crosssection}}). $\sigma_{\text{inter}}$ and
$\sigma_{\text{ano}}$ are obtained for the anomalous coupling
$\beta_0$ ($\beta_c$) in units of GeV$^{-2}$. We considered $n = 5$ and
different values of $\Lambda$; see Eq.~(\protect{\ref{ff}}).}
\label{tabcrosssection}
\end{table}


\begin{table}
\begin{tabular}{||c||c||c||c||}
Collider & Process & 
$\beta_0$ (GeV$^{-2}$) & $\beta_c$ (GeV$^{-2}$)\\
\hline 
\hline
Tevatron I & $p \bar{p} \to l^{\pm} \nu_{l^{\pm}} \gamma \gamma$ & 
( $-$4.5 , 4.4 )$\times 10^{-2}$ & ( $-$7.4 , 7.4 )$\times 10^{-2}$ \\
  & $p \bar{p} \to l^+ l^- \gamma \gamma$ & 
( $-$9.2 , 9.0 )$\times 10^{-2}$ & ( $-$15. , 15. )$\times 10^{-2}$ \\
$\Lambda$=0.5 TeV & Combined & 
( $-$4.0 , 4.0 )$\times 10^{-2}$ & ( $-$6.6 , 6.5 )$\times 10^{-2}$ \\ 
\hline
Tevatron II & $p \bar{p} \to l^{\pm} \nu_{l^{\pm}} \gamma \gamma$ & 
( $-$1.6 , 1.5 )$\times 10^{-2}$ & ( $-$2.3 , 2.2 )$\times 10^{-2}$ \\
  & $p \bar{p} \to l^+ l^- \gamma \gamma$ & 
( $-$2.9 , 2.9 )$\times 10^{-2}$ & ( $-$4.2 , 4.1 )$\times 10^{-2}$ \\
$\Lambda$=0.5 TeV & Combined & 
( $-$1.4 , 1.3 )$\times 10^{-2}$ & ( $-$2.0 , 2.0 )$\times 10^{-2}$ \\
\hline
LHC & $p p \to l^{\pm} \nu_{l^{\pm}} \gamma \gamma$ & 
( $-$2.2 , 2.1 )$\times 10^{-3}$ & ( $-$7.4 , 7.3 )$\times 10^{-4}$ \\
  & $p p \to l^+ l^- \gamma \gamma$ & 
( $-$2.4 , 2.3 )$\times 10^{-3}$ & ( $-$12. , 12. )$\times 10^{-4}$ \\
$\Lambda$=0.5 TeV & $pp \to jj\gamma\gamma$ & 
( $-$2.2 , 2.2)$\times 10^{-4}$ & ( $-$8.0 , 8.0 )$\times 10^{-4}$ \\
\hline
LHC & $p p \to l^{\pm} \nu_{l^{\pm}} \gamma \gamma$ & 
( $-$7.6 , 7.6 )$\times 10^{-5}$ & ( $-$11. , 10. )$\times 10^{-5}$ \\
  & $p p \to l^+ l^- \gamma \gamma$ & 
( $-$8.2 , 7.9 )$\times 10^{-5}$ & ( $-$14. , 13. )$\times 10^{-5}$ \\
$\Lambda$=2.5 TeV & $pp \to jj \gamma\gamma$ & 
( $-$4.4 , 4.4 )$\times 10^{-6}$ & ( $-$1.7 , 1.7 )$\times 10^{-5}$ 
\end{tabular}
\medskip
\caption{95\% CL limits on $\beta_0$ and $\beta_c$ that can be obtained at the
 Tevatron and LHC assuming that no deviation from the SM predictions is
 observed. We considered $n = 5$ and different values of $\Lambda$; see
 Eq.~(\protect{\ref{ff}}).}
\label{LFTaLHC}
\end{table}



\begin{references}


\bibitem{anomalous} For a review see: H.\ Aihara {\it et al.}, {\it
Anomalous gauge boson interactions} in Electroweak Symmetry Breaking and New
Physics at the TeV Scale, edited by T.\ Barklow, S.\ Dawson, H.\ Haber and J.\
Seigrist, (World Scientific, Singapore, 1996), p.\ 488 [hep-ph/9503425].

\bibitem{teva}
K.\ Gounder, CDF Collaboration, hep-ex/9903038;
B.\ Abbott {\it et al.}, D\O\ Collaboration,  
Phys.\ Rev.\  {\bf D62}, 052005 (2000).
  
\bibitem{lep}
R.\ Barate {\it et al.}, ALEPH Collaboration, 
Phys.\ Lett.\  {\bf B462}, 389 (1999);
P.\ Abreu {\it et al.}, DELPHI Collaboration, 
Phys.\ Lett.\  {\bf B459}, 382 (1999);
M.\ Acciarri {\em et al.\/}, L3 Collaboration, 
Phys.\ Lett.\  {\bf B467}, 171 (1999);
G.\ Abbiendi {\it et al.}, OPAL Collaboration, 
Eur.\ Phys.\ J.\  {\bf C8}, 191 (1999).

\bibitem{exp:LEP} 
G.\ Abbiendi {\it et al.}, OPAL Collaboration, 
Phys.\ Lett.\ {\bf B471}, 293 (1999); 
M.\ Acciarri {\it et al.}, L3 Collaboration, 
Phys.\ Lett.\ {\bf B478}, 39 (2000); and hep-ex/0008022.

\bibitem{lep:ichep} S.\ Spagnolo, talk at the XXXth International 
Conference on High Energy Physics, Osaka July 2000, \\
 http://ichep2000.hep.sci.osaka-u.ac.jp/scan/0727/pa05/spagnolo/.

\bibitem{hil:bes} 
A.\ Hill and J.\ J.\ van der Bij, Phys.\ Rev.\ D{\bf 36}, 3463 (1987); R.\
Casalbuoni, {\em et al.}, Nucl.\ Phys.\ {\bf B282}, 235 (1987); {\em idem}
Phys.\ Lett.\ {\bf B155}, 95 (1985); S.\ Godfrey, {\it Quartic Gauge Boson
Couplings} in Proceedings of the International Symposium on Vector Boson
Self-Interactions, edited by U.\ Baur, S.\ Errede, T.\ Muller (American
Inst. Phys., 1996), p.\ 209 [hep-ph/9505252].

\bibitem{bel:bou}
G.\ B\'elanger and F.\ Boudjema, Phys.\ Lett.\ {\bf B288}, 210 (1992).

\bibitem{ggnos}O.\ J.\ P.\ \'Eboli, M.\ B.\ Magro, P.\ G.\
Mercadante, and S.\ F.\ Novaes, Phys.\ Rev.\ {\bf D52}, 15 (1995).

\bibitem{our:vvv} O.\ J.\ P.\ \'Eboli, M.\ C.\ Gonzalez-Garcia, and S.\ 
F.\ Novaes, Nucl.\ Phys.\ {\bf B411}, 381 (1994).

\bibitem{stir} 
G.\ B\'elanger and F.\ Boudjema, Phys.\ Lett.\ {\bf B288},201 (1992); W.\ J.\
Stirling and A.\ Werthenbach, Phys.\ Lett.\ {\bf B466}, 369 (1999); W.\ J.\
Stirling and A.\ Werthenbach, Eur.\ Phys.\ J.\ {\bf C14}, 103 (2000); G.\
Belanger, F.\ Boudjema, Y.\ Kurihara, D.\ Perret-Gallix, and A.\ Semenov Eur.\
Phys.\ J.\ {\bf C13}, 283 (2000).

\bibitem{stir2} P.\ J.\ Dervan, A.\ Signer, W.\ J.\  Stirling, and
A.\ Werthenbach , J.\ Phys.\ {\bf G26}, 607 (2000).

\bibitem{bl} 
C.\ P.\ Burgess and D.\ London, Phys.\ Rev.\ {\bf D48}, 4337 (1993).

\bibitem{stu} M.\ E.\ Peskin and T.\ Takeuchi, Phys.\ Rev.\ Lett.\
 {\bf 65}, 964 (1990); Phys.\ Rev.\ D{\bf 46}, 381 (1992).

\bibitem{pdg}
D.\ E.\ Groom, {\em et al.}, Particle Data Group, Eur.\ Phys.\ J.\ {\bf C15},
1 (2000).

\bibitem{bz}U.\ Baur and D.\ Zeppenfeld, Nucl.\ Phys.\ {\bf B308}, 127 (1988).

\bibitem{mad} T.\ Stelzer and W.\ F.\ Long, Comput.\ Phys.\
Commun.\ {\bf 81}, 357 (1994).

\bibitem{helas} H.\ Murayama, I.\ Watanabe, and K.\ Hagiwara, KEK
report 91-11 (unpublished).

\bibitem{mrs} A.\ D.\ Martin, W.\ J.\ Stirling, and R.\ G.\ Roberts, Phys.\ 
Lett.\ {\bf B354}, 155 (1995).

\bibitem{vvvv} 
O.\ J.\ P.\ \'Eboli, M.\ C.\ Gonzalez--Garcia, and J.\ K.\
Mizukoshi, Phys.\ Rev.\ {\bf D58}, 034008 (1998); 
T.\ Han, H.-J.\ He, and C.-P.\ Yuan, Phys.\
Lett.\ {\bf B422}, 294 (1998);
V.\ Barger, K.\ Cheung, T.\ Han, and R.\ J.\ N.\
Philips, Phys.\ Rev.\ {\bf D52}, 3815 (1995);
E.\ E.\ Boos, H.\ J.\  He , W.\  Kilian, A.\
Pukhov, and P.\ M.\ Zerwas, Phys.\ Rev.\ {\bf D57}, 1553 (1997);
J.\ Bagger, S.\ Dawson, and G.\ Valencia,
Nucl.\ Phys.\ {\bf B399}, 364 (1993);
A.\ Dobado, D.\ Espriu, and M.\ J.\ Herrero, Z.\
Phys.\ {\bf C50}, 205 (1991); 
A.\ S.\ Belyaev, {\em et al.}, Phys.\ Rev.\ {\bf D59}, 015022 (1999);
A.\ Dobado and M.\ T.\ Urdiales, Z.\ Phys.\ {\bf C17}, 
965 (1996); J.\ R.\ Pelaez,  Phys.\ Rev.\ {\bf D55}, 4193 (1997).

\end{references}
\end{document}